\documentclass[10pt, conference, compsocconf]{IEEEtran}

\usepackage[T1]{fontenc}
\usepackage{commath}
\usepackage{amsmath}
\usepackage{multirow}
\usepackage{bm,array}
\usepackage{booktabs}
\usepackage{graphicx}
\usepackage{subcaption}
\usepackage{titling}
\newcommand*\samethanks[1][\value{footnote}]{\footnotemark[#1]}
\DeclareMathOperator*{\argmax}{argmax}
\IEEEoverridecommandlockouts
\begin{document}
%
\title{\Huge Reuters Tracer: Toward Automated News Production Using Large Scale Social Media Data}

\author{Xiaomo Liu\thanks{Two co-authors contributed equally to this work}, Armineh Nourbakhsh\samethanks[1], Quanzhi Li\thanks{Author contributed to this research when he was in Thomoson Reuters}, Sameena Shah\thanks{Corresponding Author}, Robert Martin, John Duprey\\
Research and Development, Thomson Reuters\\
\{xiaomo.liu,armineh.nourbakhsh,sameena.shah,robertd.martin,john.duprey\}@thomsonreuters.com\\
Alibaba Group Inc.\\
quanzhi.li@alibaba-inc.com\\
}
\date{}


%


\maketitle

\begin{abstract}
To deal with the sheer volume of information and gain competitive advantage, the news industry has started to explore and invest in news automation. In this paper, we present \textit{Reuters Tracer}, a system that automates end-to-end news production using Twitter data. It is capable of detecting, classifying, annotating, and disseminating news in real time for Reuters journalists without manual intervention. In contrast to other similar systems, \textit{Tracer} is topic and domain agnostic. It has a bottom-up approach to news detection, and does not rely on a predefined set of sources or subjects. Instead, it identifies emerging conversations from 12+ million tweets per day and selects those that are news-like. Then, it contextualizes each story by adding a summary and a topic to it, estimating its newsworthiness, veracity, novelty, and scope, and geotags it. Designing algorithms to generate news that meets the standards of Reuters journalists in accuracy and timeliness is quite challenging. But \textit{Tracer} is able to achieve competitive precision, recall, timeliness, and veracity on news detection and delivery. In this paper, we reveal our key algorithm designs and evaluations that helped us achieve this goal, and lessons learned along the way. 
\end{abstract}


%
\IEEEpeerreviewmaketitle

\section{Introduction}
The news industry today utilizes various communication channels to access and deliver news, from newspapers and radio, to television and internet, and more recently to mobile and social media platforms. These media outlets cooperate and also compete to deliver news to their audience in a timely fashion. Each outlet usually employs a team of journalists to collect, write and distribute news stories for specific topics. However, most outlets cannot afford to deploy their journalists around the world to cover global stories. Instead, international news agencies such as Reuters, Associated Press (AP) and Agence France-Presse (AFP) command an extensive network of journalists that cover various news topics in many countries, and provide subscription-based services to other news media. Thus, news agencies are at the forefront of news production. The newsroom of a typical news agency has a fairly standard daily workflow \cite{czarniawska2011cyberfactories}. It starts with finding both expected and unexpected events through a network of sources and channels. Some desks are dedicated to monitoring local news media, social media or collecting information through word of mouth from their personal connections. Once an event is detected, they decide whether it is worth reporting. If so, they try to verify the story and then escalate it to news editors. Editors make the final decision on the coverage and framing of the story. After a story is written, it is then distributed to media or public consumers through the agency's wire service. From this workflow, we can observe that global news agencies like Reuters have to cover all topics of impactful news with high speed and accaurcy in order to thrive in the news business. 

The advent of internet and the subsequent information explosion has made it increasingly challenging for journalists to produce news accurately and swiftly. Some agencies have begun to computerize parts of their workflow. They employ algorithms to tailor news stories to different customers' needs \cite{wang2015nyt}. They also use software to compose certain types of stories automatically \cite{colford2014leap}. It is commonly agreed that automation is critical to the future of news agencies. Automation can help news outlets detect breaking stories quickly and give them a competitive advantage in timeliness \cite{Stray:2016}. In light of the ``fake news'' controversy and the public debate around media bias and credibility, automated verification of news has also become an important issue. Detecting news from a universal pool of tweets can help editors avoid the ``echo-chamber effect'' and an automated verification algorithm can at least partially alleviate the burden of manual, case-by-case vetting of stories.

In this paper, we present a news automation system, called \textit{Reuters Tracer}, currently deployed at Reuters News Agency \cite{Al-Kofahi:2017}. Unlike other systems, this product automates each aspect of news production: from finding news, to classification, annotation, storytelling, and finally to dissemination \cite{diakopoulos2016algorithmic}. There are many challenges to inventing and productizing this system, from both technical and product perspectives. First, journalism is a profession that requires specialized training and skills. Journalists have complex daily workflows and are thus indispensable when it comes to automation. We have created 10+ machine learning algorithms that work together in a unified system that can automate the entire news production pipeline of a news agency, to our knowledge, for the first time. Second, the automated news production system needs to reach the professional level of reporting speed and accuracy as well as news coverage in order to be valuable to news agencies since journalists will never sacrifice these core news values in exchange for less work. Our algorithm and system design are proritized to sustain or even enhance these core values. \textit{Tracer} utilizes Twitter data and processes 12+ million tweets every day. It can reduce 99\% of ingested data, mostly noises, and distill news that covers 70\% of daily news reported by journalists of global news media and agenicies such as Reuers, AP and CNN (can get 95\% if using 50+ million tweets).  We deployed \textit{Tracer} on a big-data processing infrastructure with a cluster of 13 nodes. It can guarantee the end-to-end system latency, from consuming tweets to produce news, is within about 40 milliseconds scaling up to 50+ million tweets. With suites of algorithms invented by us, \textit{Tracer} is able to detect news with high precision \& recall and, most importantly, faster than the army of Reuters' news professionals. Our benchmark results illustrate it has beaten global news outlets in breaking over 50 major news stories and given Reuters journalists an 8 to 60-minute head-start over other news media. \textit{Tracer} relies on Twitter to gain speed advantages, but also faces the risk of rampant misinformation on social media. To tackle this problem, we designed an algorithm to estimate the veracity of detected news. Our experiments show it can detect online rumors and fake news with 60-70\% accuracy when they emerge as events in \textit{Tracer} system and keep improving the estimation capability as more tweets are captured. Therefore, this work not only brings significant business value to Reuters and its customers, but also advances the journalism technology in general. In the rest of paper, we reveal how each of algorithms and overall system are designed to achieve these outcomes.

\section{Related Work}
Finding, making sense of, telling, and disseminating news stories are the main procedural aspects of news production conducted by human journalists \cite{diakopoulos2016algorithmic}. As Zuboff's law - \textit{everything that can be automated will be automated} - makes its mark on journalism, each of these four aspects is undergoing automation. Systems like \textit{Social Sensor} have been invented to search on social media and identify ongoing trends for finding news \cite{schifferes2014identifying}. After information is collected, making sense of stories from the raw text stream is a daunting challenge. A system that overcomes this difficulty is \textit{TwitInfo} \cite{marcus2011twitinfo}. It employs a suite of algorithms and visualizations to analyze the wax and wane of long-run events, the crowds' sentiment around sub-events, and their geographic footprints. Despite advances in news automation, fully machine-driven storytelling is far from achieved. But news industry has made significant progress in this regard. Most notably, Associated Press (AP) partnered with a software vendor to automatically generate thousands of reports of specific topics such as sports match summaries and quarterly earnings \cite{colford2014leap}. When new stories are about to be disseminated, algorithms can again play an important role. For example, \textit{The New York Times} built a bot to help editors select a limited number of featured stories each day in order to maximize the audience \cite{wang2015nyt}.      

Although all of these efforts advanced news automation in various aspects, none of them were able to computerize the detection, contextualization, narrativization and dissemination of news in a unified system. To fully automate news production, these four aspects need to be powered by intelligent algorithms. The existing solutions closest to this goal are the natural language news generation systems operating in AP, Yahoo! News, etc \cite{diakopoulos2016algorithmic}. The underlying technology is based on structured data and pre-written templates. For example, earning reports published by AP follow a standard format such as ``\textit{X reported profit of Y million in Q3, results beat Wall Street forecasts...}". The company name, its profit figure and the forecast come from data providers. The template is constructed by rule-based sentence selection to express different scenarios and the language is enriched by dynamic synonyms. This technology is best suited to producing routine news stories of repetitive topics \cite{graefe2016guide}. Unfortunately, the majority of important news stories are emergent and cannot be automated by this method. Besides, there are also several systems like \textit{TwitterStand} \cite{Sankaranarayanan:2009} for finding news from social media. But most of them use much smaller data streams comparing to us and are not designed to detect news before they break into mainstream media. In this regard, the existing technologies are far from being called an ``automated news agency". In the rest of this paper, we describe how our system was architected to achieve this goal.

\begin{figure}
\centering
\includegraphics[width=0.45\textwidth]{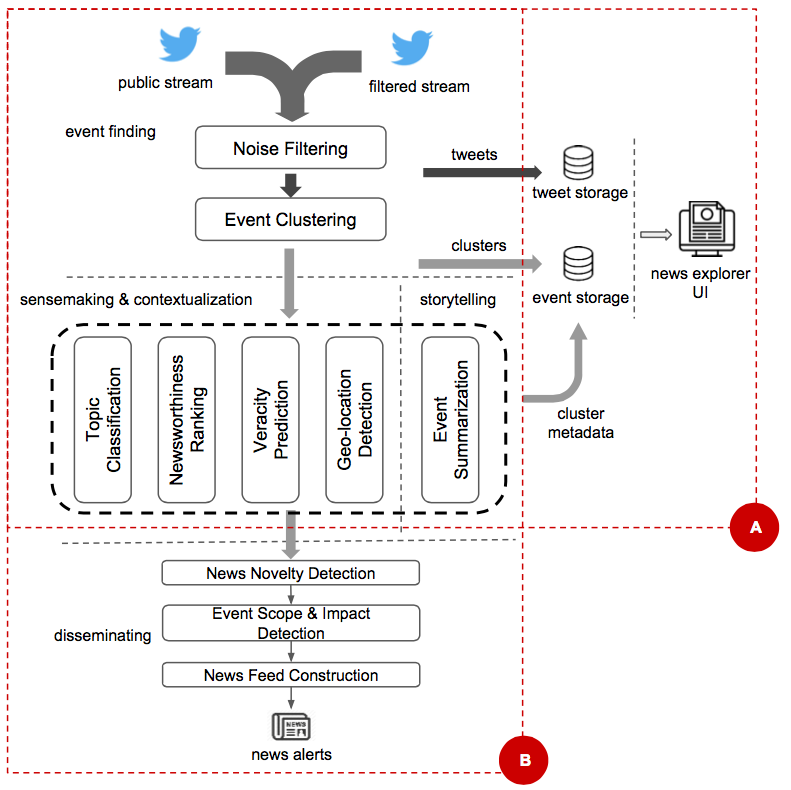}
\caption{Tracer's system architecture for two use cases: (A) news exploration UI; (B) automated news feeds.} 
\label{fig:tracer_arch}
\end{figure}

\section{Data Source}
We start with our rationale behind choosing Twitter as the data source for \textit{Tracer}. Social media gives immense power to citizen journalists and eyewitnesses to spread information in case of events. Internal research by Reuters found that 10-20\% of news breaks first on Twitter. This observation was confirmed by a study showing that social media, especially Twitter, can lead newswires in various topics \cite{osborne2014facebook}. Thus, we decided to use Twitter streams to cultivate its speed advantage. \textit{Tracer} uses 2\% of Twitter data (about 12+ million tweets everyday) through streaming APIs: (1) a pubic stream\footnote{https://dev.twitter.com/streaming/reference/get/statuses/sample} that is a random sample of 1\% of all public tweets; (2) a filtered stream\footnote{https://dev.twitter.com/streaming/reference/post/statuses/filter} that is filtered by a taxonomy of news-related terms and users curated by journalists. A global team of Reuters journalists from American, European, Asian and African desks contributed to the creation of this list. Their lists were collected and ranked by importance and popularity. Then 376 key words and 4,993 user accounts were selected to create the filtered stream, which the Twitter API bounds to 1\% of total Twitter traffic. As our evaluation in Section \ref{sec::perf} shows, these two streams helped \textit{Tracer} recall almost 70\% of headlines published by Reuters, AP and CNN and achieve significant speed advantages compared to mainstream news media. This means that the system's coverage and timeliness are competitive, despite its limited, but still large scale, use of Twitter data.

\section{Algorithm Suites}
The core algorithms of \textit{Tracer} are organized into four suites, each implementing one aspect of news production.

\subsection{Event Discovery}
Although Twitter is an excellent source for monitoring events, it submerges useful information in a flurry of noise. Fishing out breaking stories is a hard technical challenge. In our prior publication, two algorithms - noise filtering and event clustering - were proposed to tackle this problem \cite{liu2016reuters}. Herein, we only briefly describe the general ideas behind these two algorithms and focus on the remaining three downstream algorithm suites, which are more critical to news automation.

\textbf{Noise Filtering}. We first conceptualize the Twitter stream as a spectrum of information from noise to news, which consists of spam, advertisements, chit-chat, general information, events, normal news, and breaking news. The task of filtering noise is divided and conquered by sub-algorithms that distinguish each type of noise out of events and news iteratively. This strategy helps alleviate the problem of unbalanced data, which often misguides classifiers. To ensure no stories are missed, the algorithm is tuned to penalize misclassifying non-noise, i.e. false negatives. This design tolerates some level of noise and ensures that more events can be detected. Our experiment in Table \ref{tbl:news_metric} reveals that more than 78\% tweets can be filtered after this step with less than 1\% false negatives \cite{liu2016reuters}.

\textbf{Event Clustering}. Like most prior studies, we consider news finding in Twitter as a problem of event detection via clustering tweets \cite{Atefeh2013}. This design assumes that if a group of people talk about the same subject at a particular time, it is likely to be an event. Unlike most traditional methods, our clustering algorithm consists of two phases: clustering and merging. The cluster phase first produces unit clusters of 3 tweets with similar content, after which they are merged with a pool of existing clusters. As long as a unit cluster forms and doesn't result a merge, it is called out as an event. This design speeds up detection and simplifies periodical cluster updates \cite{Yin:2013}. We benchmarked our algorithm with a locality sensitive hashing (LSH) based event detection algorithm recently proposed to handle big data \cite{petrovic2010streaming}. Our algorithm can find more events, because LSH is an approximate search algorithm, whereas our algorithm can operate full search supported by our big data processing infrastructure (see Section \ref{sec:data_system}).

\subsection{Sensemaking \& Contextualization}
Once emergent events are collected, journalists need to process, organize and make sense of them in order to come up with newsworthy stories. To emulate this process, \textit{Tracer} employs a suite of algorithms to capture the contexts of detected events. These algorithms identify: the topic of the event; where it happened; whether it is newsworthy; if it seems like a true story or a false rumor. We've reported our event topic classification algorithm that uses semantic enrichments beyond \emph{TD-IDF} by incoroprating by distributed language representation \cite{li2016tweet}. This idea is not alone. Many other studies have considered additional langauge semantics to impove their classification tasks, such as \cite{chen2010semantic}. In this paper, we focus the discussions on our newsworthiness and veracity algorithms. 

\textbf{Newsworthiness Detection}.
Newsworthiness is formally defined by journalists via several factors: topic of human interest, presence of prominent subjects (people, organizations and places), public attention and personal perception \cite{harcup2001news,freitas2016identifying}. Therefore, our algorithm captures the subjective aspects of newsworthiness that mainly depend on the content of an event. Newsworthiness is also a dynamic concept, since news topics can shift and emerge over time \cite{Leskovec2009meme}. We also consider the news dynamics by modeling short and long term newsworthiness. Based on these requirements, we define the problem of newsworthiness detection as follows. Given an event cluster $e$ and its associated tweets $\{w_{1},...,w_{m}\}$, our algorithm should predict its probability of being news as $p(e)$, which aggregates the probabilities of having news topics $p_{T}(e)$, containing news objects $p_{O}(e)$, and attracting public attention $p_{A}(e)$.  

\textit{Model of News Topics}. Finding newsworthy topics from Twitter stream in practice cannot be solved as a text classification problem. The amount of news is relatively small and the number of news topics may have a limit. But non-news topics such as mundane conversations on social media are much larger and change on a daily basis. Thus, there is no economical way to collect a dataset with statistically-representative samples of non-news. Our strategy is to design an algorithm that works like a one-class classifier \cite{khan2009survey}. The rationale is that if we can collect a high quality set of newsworthy tweets through news media accounts on Twitter, we can define a reasonable decision boundary around these positive samples. We first identified 31 official news accounts on Twitter including news agencies like \texttt{@Reuters} and \texttt{@AP}, news outlets like \texttt{@CNN}, \texttt{@BBCBreaking} and \texttt{@nytimes}, and news aggregators like \texttt{@BreakingNews}. Their tweets over a one-year period (2014.8-2015.8) constitute the positive dataset. We then use the dataset to train a news topic model $Z_{l}=\{z_{1},...,z_{n}\}$ using Gibbs sampling with $n$ topics, where $n$ is determined by minimizing the model perplexity \cite{griffiths2004finding}. This model is used to infer topic probabilities of cluster tweets. If a tweet is not newsworthy, the sampler tends to assign it to a random topic uniformly and results in a flat topic distribution. If it is about news, then the inferred distribution is skewed to certain topics. Thus, we calculate news probability of a cluster $e$ as $P_{T_{l}}(e) = \sum_{i} \sum_{t5} p( Z_{l} | w_{i} )/ m$, where $\sum_{t5} p(Z_{l}|w_{i})$ is top 5 (determined by experiment) topic probabilities of $w_{i}$. The overall cluster probability is an average on all tweets in the clusters.  \textit{Model of News Object} uses the same news dataset. We extract all of name entities $S=\{s_{1},...,s_{k}\}$ and compute their frequency distribution as news probabilities. The probability of a tweet $w_{i}$ in this model is then $p_{O_{l}}(w_{i})= \sum_{t2}p(s_{i}|w_{i})$, where $\sum_{t2}p(s_{i}|w_{i})$ is its top 2 news probabilities\footnote{News usually involves 1 or 2 objects like people and places.}. Again, the overall cluster probability averages on its tweets. Besides long term models of news topics and objects, we also build the corresponding short term models $p_{T_{s}}$ and $p_{O_{s}}$ using the preceding month's news dataset. \textit{Model of Public Attention} measures the impact of cluster size (as a measure of public engagement) to newsworthiness. We simply compute its probability as $p_{A}(e) = \log_{10}\norm{e}/\log_{10}S$, where $\norm{e}$ is cluster size and $S$ is a predefined cluster size limit. Based on our observations, we set $S=500$ which sets $p_{A}(e)=1$ if cluster $e$ has 500+ tweets. The final newsworthiness score is learned by ordinal regression which suits our multi-grade newsworthiness data (see Section \ref{subsec:news}).

\textbf{Veracity Prediction}. Social media has become an important information supplier for news media, but it also contains a flurry of misinformation, rumors, and even fake news. In the course of 2016 US president election, Facebook was fiercely criticized for not being able to combat widespread fake news that favored one candidate \cite{SilvermanFakeNewsFacebook}. With this challenge in mind, we designed an algorithm for estimating the veracity of news detected by \textit{Tracer} to keep our users aware of potential risks. We have explained our verification algorithm as a standalone application in \cite{liu2015rumor}. Here we disclose how it works with other components in \textit{Tracer} on streaming data. Our lesson learned is that credibility \cite{castillo2011information} does not necessarily equal veracity since even celebrities and news media are occasionally fooled by false rumors \cite{nourbakhsh2015rumor}. Credibility is just one factor used by journalists to check news veracity \cite{silverman2014verification}. They also care about who originates the news, if the source is credible and has legitimate identity, if there are multiple independent sources, etc. Thus, we encode these verification steps into our algorithm to make it more efficient. Specifically, we train multiple SVM regression models with different features to operate on early and developing stages of an event separately. These models produce scores in $[-1,1]$ to indicate the degree of veracity.

\textit{Early Verification}. When a rumor is just originated and not yet widespread on social media, it is likely captured by \textit{Tracer} as a small-sized event cluster. The challenge here is to make a reliable prediction as to whether the rumor is true, with very limited information. Journalists often use the tweets as well as the sources that they cite to identify the veracity of a rumor at this stage. Our algorithm identifies the source using three rules: (1) if an event tweet is a retweet, the original tweet is its source; (2) if it cites a url, the cited webpage is the source; (3) the algorithm issues a set of queries to the Twitter search API to find out the earliest tweet mentioning the event. The credibility and identity of the source are good indicators of event veracity. For example, if it turns out the information source is from a satirical news site such as \texttt{The Onion} or a fake news site like the \texttt{National Report}\footnote{nationalreport.net}, then our algorithm will likely flag this event as false.

\textit{Developing Verification}. Once an event gains momentum, its \textit{Tracer} cluster collects more tweets. At this stage, public reaction to the event provides additional context to our algorithm. There are often a few people who express skepticism about false rumors or even directly debunk them \cite{zhao2015enquiring, liu2015rumor}. They provide useful clues to the verification algorithm. We built a sub-algorithm in \textit{Tracer} to identify people's expressions of belief, including negation (e.g. ``this is a hoax"), question (e.g. ``is this real?"), support (e.g. ``just confirmed") and neutrality (mostly retweets). Hence, we conceptualize the veracity prediction task as a ``debate" between two sides. Whichever side is more credible wins the debate.

\subsection{Storytelling}
Unlike other algorithmic storytelling systems that can generate stories from structured data \cite{graefe2016guide}, \textit{Tracer}'s documents (tweets) are unstructured and short, with fewer than 140 characters. Therefore, instead of generating full stories, the system generates a short headline for each event. This design also aligns with Reuters' internal alerting system\footnote{https://goo.gl/Vp3vS8}, where breaking stories are broadcast internally as short headlines. Since \textit{Tracer} detects events as clusters, then cluster summary is the most straightforward choice of a news headline. 

\textbf{Event Summarization} Besides representativeness, the selected summary of each event needs also be readable and objective to meet the standards of news headlines. For example, ``BREAKING: Donald Trump is elected president of the United States" is a preferred summary over ``OMG! Trump just won!!! \#Trump4Presdent". Because the latter contains personal emotions, short-hand notations and misspellings. We treat this task as selecting an appropriate tweet from tweets $\mathcal{E}$ in an event cluster. One of most widely used text summarization algorithms is \texttt{LexRank} \cite{erkan2004lexrank}. As pointed out by Becker et al. \cite{becker2011selecting}, however, \texttt{LexRank} doesn't work well on tweet clusters as it strongly favors repeated language (retweets) and diverges from the main event topic. Therefore, our algorithm selects event summaries based on the cluster centroid while penalizing incorrect and informal language. Each tweet in $\mathcal{E}$ is converted to a vector $\overrightarrow{w_{i}}$ of \textit{tf-idf} representation. The most representative tweet is closest to the centroid $\overrightarrow{C_{\mathcal{E}}}$. The centroid helps avoid low quality text since \textit{tf} highlights important terms and the penalty term improves the readability and objectivity of summaries as follows,  
\begin{equation}
	w_{i} = \argmax_{w_{i} \in C_{\mathcal{E}}}(\mbox{sim}(\overrightarrow{w_{i}}, \overrightarrow{C_{\mathcal{E}}}) - \lambda \mathcal{I}(w_{i}))
\end{equation}
Parameter $\lambda$ controls the strength of penalty. The $\mathcal{I}(w_{i})$ is an indicator function for rule-based informal language detection in a tweet. The rules capture indicators such as the presence of out of vocabulary words (except named entities), hashtags in the middle of a tweet, Twitter handles that don't belong to orgs or high-profile users, and 1st \& 2nd person pronouns.

\subsection{Dissemination}\label{dissemination}
\textit{Tracer}'s algorithms covered so far are designed to cater to the needs of a typical Reuters journalist. They detect news stories across all newsworthy topics (from a movie award to a terror attack), and report events of various scopes (from a small building fire to a large forest fire). However, news consumers have different interests based on their professional or personal needs. The last step of news automation is to select news alerts (or headlines) tailored to diverse customer requirements. Thankfully \textit{Tracer}'s pipeline is flexible, and its algorithms are parameterized in such a way that makes them easy to customize. In this section, we present algorithms that help us customize the \textit{Tracer} feed along three dimensions: \textit{novelty}, \textit{scope/impact}, and \textit{localization}.

\textbf{Novelty} refers to the recency of a story. \textit{Tracer} detects new (i.e. ``breaking'') stories as well as updates and follow-ups. For instance if a terror attack takes place, the standard \textit{Tracer} feed will report it as a headline. If the number of casualties is confirmed, it will be reported as a separate headline. If a suspect is arrested, it will be yet another headline, and so on. Occasionally, social media users discuss events that may be as old as days, months, or even years. For instance, they might commemorate events from World War II. A real-time event detection system should only report events that have happened very recently. As a requirement provided by end users, we define stories that have taken place more than 12 hours ago as old or ``stale'' events.

To filter out old events, we employ a hybrid approach that uses two indicators: 1) the temporal information expressed in the tweets, and 2) the previous history of events reported by the system. If an event has expressions of staleness such as ``yesterday,'' ``last week,'' ``in 1945,'' or ``\#onthisday,'' it will be discarded as old. If a tweet includes temporal expressions such as ``on Monday,'' or ``this morning,'' then its timestamp is used to determine whether the event it is discussing has occurred more than 12 hours ago. If an event does not include any such expression but is too similar to a previously reported event, it will also be discarded. Similarity is measured as the distance between six pairs of vectors, representing the semantic dimensions of an event (``who,'' ``what,'' ``where,'' ``when''). A cut-off threshold is learned to determine when an event should be considered too similar to a previously reported headline. The details of this algorithm are discussed in \cite{li2017novelty}.

\textbf{Scope/Impact} refers to the potential extent of an event's impact, including its magnitude (such as the scale of an earthquake), human impact (such as casualties or injuries) or financial/physical impact (such as damage to infrastructure, to a business, or to a nation's economy). As mentioned previously \textit{Tracer} assigns a topic to each event. Most topics (such as \textit{Entertainment} or \textit{Sports}) have a tendency to have flexible or undefined scopes, but three topics (\textit{Crisis}, \textit{Law/Crime} and \textit{Weather}) carry a specific sense of scope depending on the type of event they are reporting. For instance an apartment fire has a small scope and might not be relevant to a disaster journalist, but an explosion at an oil pipeline is. We employ a cardinal-extractor and a linear classifier to identify impact in natural and man-made disasters. The details of the algorithm have been discussed in \cite{nourbakhsh2017disasters}

\textbf{Localization} refers to the need to place each event on a global map. Many users of \textit{Tracer} are local journalists, or businesses with certain geographic areas of interest. The tool would not be of use to them if they were not able to filter their feed for a particular area or location. Thus, we created a geo-parsing and geo-coding suite that places each event along latitudinal and longitudinal dimensions. Both geo-parsing and geo-coding of microblogs are complicated information retrieval tasks that have attracted a large body of research \cite{Ozdikis:2016}. We design our tweet-based geo-parsing model on a hybrid system \footnote{Details will be disclosed in upcoming publications.} that uses different methodologies for different types of geo-locations. It uses a small taxonomy of 3,239 location names to detect large geolocations including continents, countries, and large states and provinces. To detect cities, it combines a heuristic method inspired by the schema introduced in \cite{Mandl2008} with a rank-reciprocity algorithm that helps it detect misspellings (e.g. ``illinoi''), abbreviations (``pdx''), hashtags (``\#okwx''), and aliases (``Nola''). 

The resulting toponyms are sent to a geo-coding service \footnote{http://wiki.openstreetmap.org/wiki/Nominatim} and mapped to latitude/longitude dimensions. This service creates multiple candidate lat/lon pairs for each toponym. For instance, ``Paris'' can be mapped to \textit{Paris, France} or \textit{Paris, Texas}. Our model uses user profile information and other event metadata to pick one candidate that is likely to match the true location of the toponym.

\begin{figure}
\centering
\includegraphics[width=0.48\textwidth]{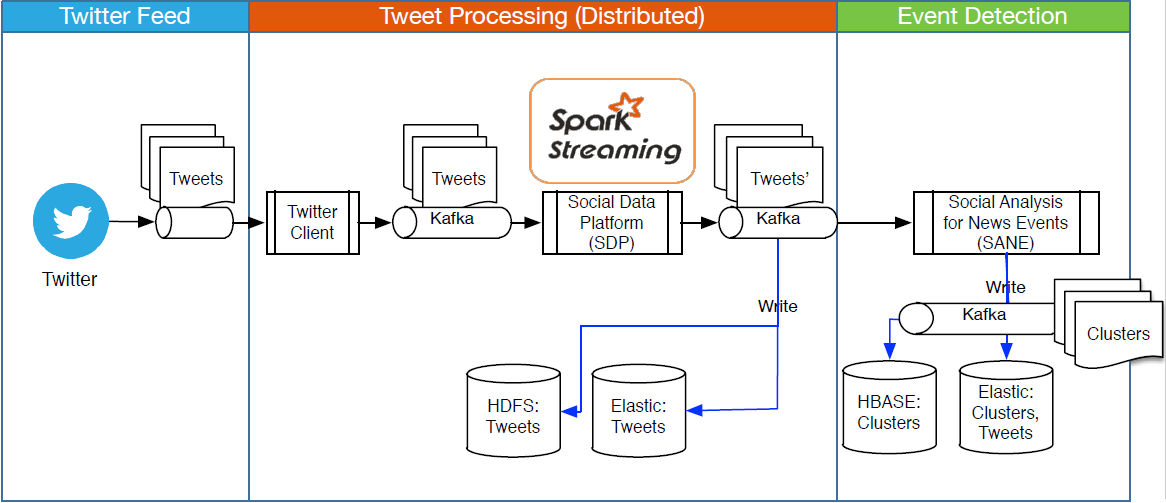}
\caption{Data processing system architecture for \textit{Tracer}.} \label{fig:sdp_arch}
\end{figure}

\section{Data Processing Infrastructure} \label{sec:data_system}
The above algorithms resolve the problem of how to detect, verify and deliver news from massive data. To facilitate them to process 12 million tweets per day with low latency, we also engineered a big data infrastructure for \textit{Tracer}, consisting of a cluster with 13 nodes at our data-center. Figure \ref{fig:sdp_arch} presents the overall architecture of this infrastructure.

\textbf{Data Processing}. The whole system scales through the use of durable and resilient Kafka message topics and the use of horizontal scalability of Spark Streaming. Tweets ingested from Twitter stream are first written to a Kafka topic. The downstream pipeline subscribes to this topic and implements a series of Spark streaming jobs to run tweet-level machine learning models such as language detection, entity extraction and noise filtering. The tweets along with their model outputs are written to HDFS and Elastic indexes for future analysis. We named this process the \textit{Social Data Platform (SDP)}. Event detection processing, which is called \textit{Social Analysis for News Event (SANE)}, is similar to SDP's distributed processing of tweets. Unlike SDP, it uses Apache Camel to orchestrate invocation of the analyzers. Both inputs and outputs of event detection use Kafka again to avoid an I/O bottleneck. Our benchmarking result shows that it successfully processes more than 12 million tweets every day with a 38.4 millisecond latency on average, 6 millisecond median. 

\textbf{News Delivery}. When delivering the detected clusters to end users, clusters are output to another Kafka topic where they can be persisted to HBase and pushed to an in-memory database for the fast access by web UI. A separate process consumes the cluster data in HBase and executes the novelty, scope and localization models in parallel to produce multiple instances of \textit{Tracer} news feeds.  

\begin{figure*}
\centering
\includegraphics[width=0.9\textwidth]{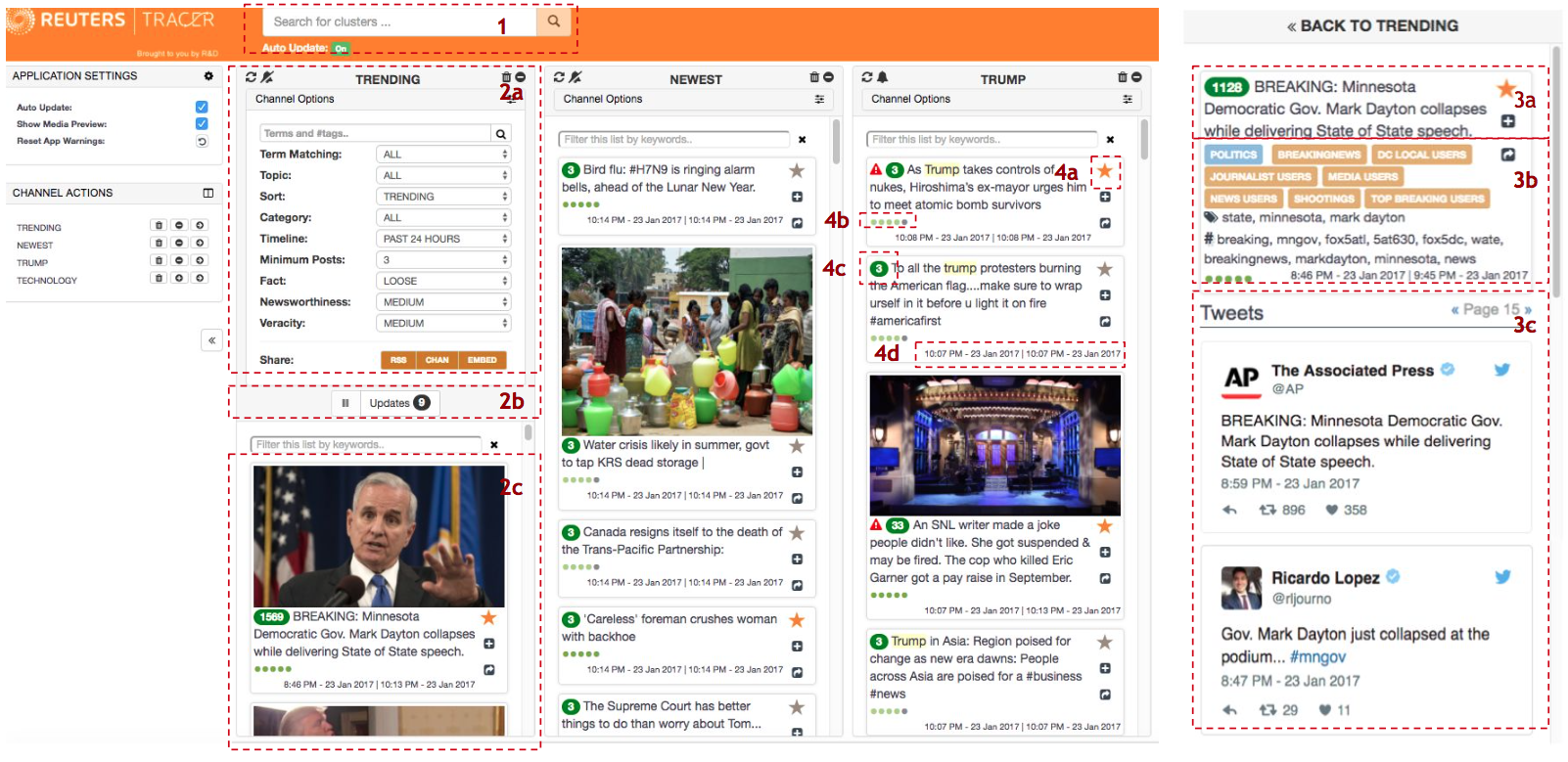}
\caption{Tracer's news exploration UI. \textbf{(1)} Global search; \textbf{(2)} News channel (2c) with editable channel options (2a) and live updates (2b); \textbf{(3)} News cluster with its summary (3a) and metadata (3b) as well as its associated tweets (3b); \textbf{(4)} Cluster metadata including newsworthy indicator (4a), veracity indicator (4b), cluster size (4c), and created \& updated times (4d).} 
\label{fig:tracer_ui}
\end{figure*}

\section{News Automation Platforms}
\subsection{Semi-Automation: Web UI for Tracer}
After \textit{Tracer} runs on the above algorithms and infrastructure, it generates a set of enriched event clusters. We built a web service for journalists to explore these clusters and pick and choose events that might be relevant to them. This use case is a semi-automated application of \textit{Tracer} in the news room, since it relies on the discretion of journalists. In this sense, it positions \textit{Tracer} as a highly intelligent social monitoring tool. Figure \ref{fig:tracer_ui} displays a screenshot of this web interface.

The \textbf{Search Box} provides Google-style search and tracking capability. Users can look-up queries such as ``US election''. A news tracking channel will be created with clusters matching the query terms. Boolean operators are provided via the ``Term Matching'' option. Hashtags, cashtags and exact quotes are also recognized. Queries are not mandatory; a channel can be created with any filter on the Cluster Options panel (Figure \ref{fig:tracer_ui}, 2a). Users can create multiple channels to track different topics. Each \textbf{News Channel} is live, i.e. it not only retrieves historical clusters but also pops up newly detected or updated clusters to the top. To further refine the channel, it has options to tune its filtering criteria by newsworthiness, veracity, topic, cluster size, and time span. If a user clicks on a cluster, a \textbf{Cluster View} displays details of a cluster including all of its tweets. Cluster tweets are sorted in reverse chronological order, and retweets are de-duplicated to reduce redundancy. For each cluster, a set of \textbf{Cluster Metadata} is displayed, including a summary, a newsworthiness indicator which tells whether the event is at least partially newsworthy, and a 5-dot\footnote{Currently, we evenly divide $[-1,1]$ veracity scores into 5 dots. A better method needs to be proposed in future.} veracity indicator which predicts true (4-5 green dots), suspicious (3 yellow dots), and false (1-2 red dots) news. The metadata helps journalists to make sense of events quickly and conveniently.

\subsection{Full-Automation: Eikon Feed for Tracer}\label{sc}
\begin{figure}
\centering
\includegraphics[width=0.5\textwidth]{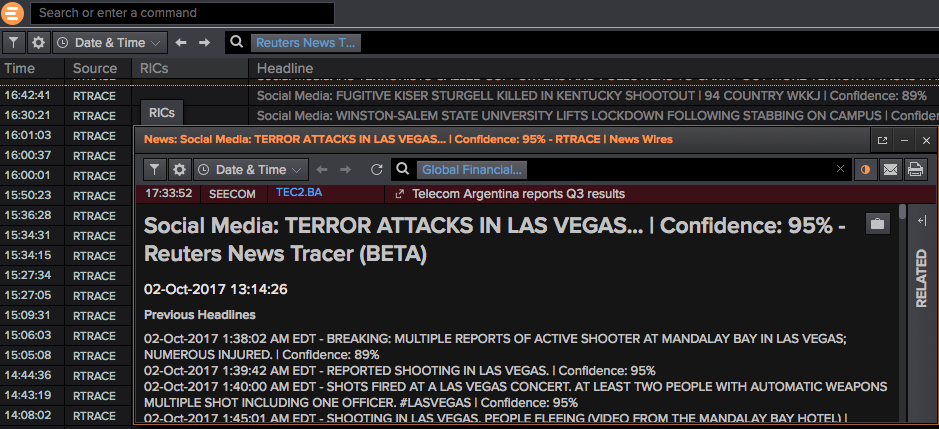}
\caption{A live Tracer news feed (ticker name TRACE) in Eikon.} 
\label{fig:TRACE_feed}
\end{figure}

The GUI above provides a seamless way of accessing news for journalistic and personal purposes. However, for some business cases and instantaneous news delivery, it is more appropriate to have a fully-automated alerting system to publish news without any human intervention. Here we describe two such cases for which \textit{Tracer} was transformed into a fully automated news feed (see Figure \ref{fig:TRACE_feed}), provided to clients via the Eikon\footnote{https://financial.thomsonreuters.com/en/products/tools-applications/trading-investment-tools/eikon-trading-software.html}, Thomson Reuters' information monitoring and analytic platform for financial professionals. 

\textbf{Disaster Feed:} A common category of ``breaking'' stories are unexpected disasters such as manmade or natural disasters, accidents and sudden outbreaks of social unrest. Disasters are the second most prominent topic that breaks on Twitter ahead of traditional news media \cite{Macdonald:2013}. Their business value goes beyond journalism and appeals to various financial and legal practices, from risk profiling to investment management. We have created a custom feed of news events using \textit{Tracer}'s pipeline, but focusing specifically on disasters. To achieve this, we filtered \textit{Tracer}'s events down to those for which the \textit{Topic} was either \textit{Crisis}, \textit{Crime}, or \textit{Weather}, \textit{Novelty} was \textit{high}, \textit{Scope/Impact} was \textit{medium} to \textit{high}, and granular geo information was available. Granularity of localization is defined per event type. For instance the largest acceptable locality for a terror attack is a city, but floods can cover large areas and multiple provinces. The resulting feed generates an average of 101 headlines on a daily basis. Table \ref{tab-superfeeds} shows a few examples of headlines and metadata generated by the disaster feed.  

\textbf{Supply Chain Event Feed:} Acts of terrorism are one of the top 10 causes of disruptions in the supply chains of businesses. Other physical events such as natural disasters, accidents and fires make up a large portion of supply chain disruptions \cite{bci:2016}. Detecting these events in a timely manner helps businesses plan and strategize accordingly. To adapt \textit{Tracer}'s feed to detect supply chain events, we simply tuned the parameters mentioned above. First, we lowered the \textit{Scope/Impact} parameter. This means that small fires and low-impact natural disasters were not removed, since they can disrupt supply chains. Instead, we tightened the localization parameter. Many business are only interested in areas where their suppliers or transportation channels are located. By making the localization parameter more granular, we made sure that that events were pin-pointed to the smallest available level of granularity. For example, in the disaster feed, we reported fires at the city or district level, but for the supply chain feed we limited the system to report fires at the best available landmark or address.

\begin{table}[]
\centering
\caption{Three sample headlines generated by \textit{Tracer}'s automated disaster feed (compared with earliest Reuters alerts).}
\label{tab-superfeeds}
\resizebox{\columnwidth}{!}{%
\begin{tabular}{l|l|l|l}
\hline
\textbf{Date} & \textbf{\begin{tabular}[c]{@{}l@{}} Tracer \end{tabular}} & \textbf{Tracer Headline}                                                                                                                                                          & \textbf{\begin{tabular}[c]{@{}l@{}} Reuters alert \end{tabular}} \\ \hline
2016-09-01    & 09:27:12                                                           & \begin{tabular}[c]{@{}l@{}}Awaiting verification and more\\ news about a reported explosion \\ at @SpaceX facility in Cape Canaveral.\end{tabular}                         & 09:35:26                                                                           \\ \hline
2016-10-28    & 15:46:20                                                           & \begin{tabular}[c]{@{}l@{}}Fire at O'Hare International Airport in \\ Chicago \#ORD\end{tabular}                                                                           & 16:09:32                                                                           \\ \hline
2016-12-31    & 18:03:18                                                           & \begin{tabular}[c]{@{}l@{}}Breaking: shooting at Turkish nightclub\\ in Istanbul. Many injured reported so far. \\ No details yet, possible terrorist attack.\end{tabular} & 18:06:37                                                                          
\end{tabular}%
}
\end{table}

\section{Tracer's Performance Analysis} \label{sec::perf}
In this section, we conduct an extensive analysis of \textit{Tracer}'s performance in terms of how many and how accurately and quickly it can detect news in real time. 

\subsection{Event Detection}  \label{subsec:coverage}

\begin{table}
\centering
\caption{Statistics of tweets processed and events detected by \textit{Tracer} in the week of 8/11-8/17, 2017 (compared with Reuters journalists).} \label{tbl:news_metric}
\setlength\tabcolsep{5pt}
\begin{tabular}{|c|c|c|c|c|c|c|c|} \hline          
        & \multicolumn{3}{c|}{\textbf{Tweets}}  & \multicolumn{2}{c|}{\textbf{Clusters}} & \multicolumn{2}{c|}{\textbf{Reuters}} \\ \cline{2-8} 
         &  All       &  Non-Noise & Clustered  & All    & News.   &  Alerts & News        \\ \hline \hline
Daily    & 12M        & 2.6M       & 624,586    & 16,261    & 6,695   & 3,360   & 255         \\ \hline
Hourly   & 512,798    & 107,233    & 26,024     & 678       & 279     & 140     & 11          \\ \hline
\end{tabular}
\end{table}

\textit{Tracer}'s one week usage data in Table \ref{tbl:news_metric} shows that it processes 12+ million tweets each day, of which 78\% are filtered as noise. In the subsequent clustering stage, only 5\% of tweets are finally preserved to produce 16,000+ daily event clusters on average. Our algorithm is able to further identify 6,600+ events that are potentially newsworthy. These are all achieved by an automated system with a a cluster of 13 servers and 10+ machine learning algorithms. In contrast, Reuters nowadays deploys 2,500+ journalists across 200+ world-wide locations. They bring back 3000+ news alerts to the internal event notification system every day. Among those, 250+ events on average are written as news stories and broadcast to the public. Even though \textit{Tracer} uses only 2\% Twitter data, it can detect significantly more events than news professionals. We further conducted a study of the news coverage by \textit{Tracer}. A set of 2,536 news headlines from Reuters, AP and CNN in a week from 05/08/2016 was selected and compared to events detected by \textit{Tracer}. The results indicate \textit{Tracer} can cover about 70\% news stories with 2\% free Twitter data and increase to 95\% if we pay to obtain 10\% data.

\subsection{Newsworthiness} \label{subsec:news}

\begin{figure}
\begin{subfigure}{.5\textwidth}
  \centering
  \includegraphics[width=1\linewidth]{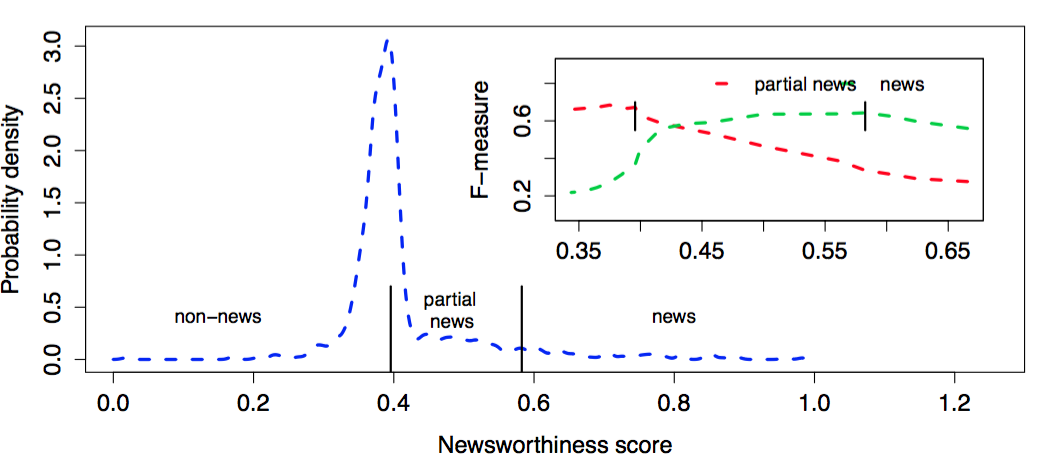}
\end{subfigure}\\[1ex]
\begin{subfigure}{.5\textwidth}
  \centering
  \includegraphics[width=1\linewidth,scale=1]{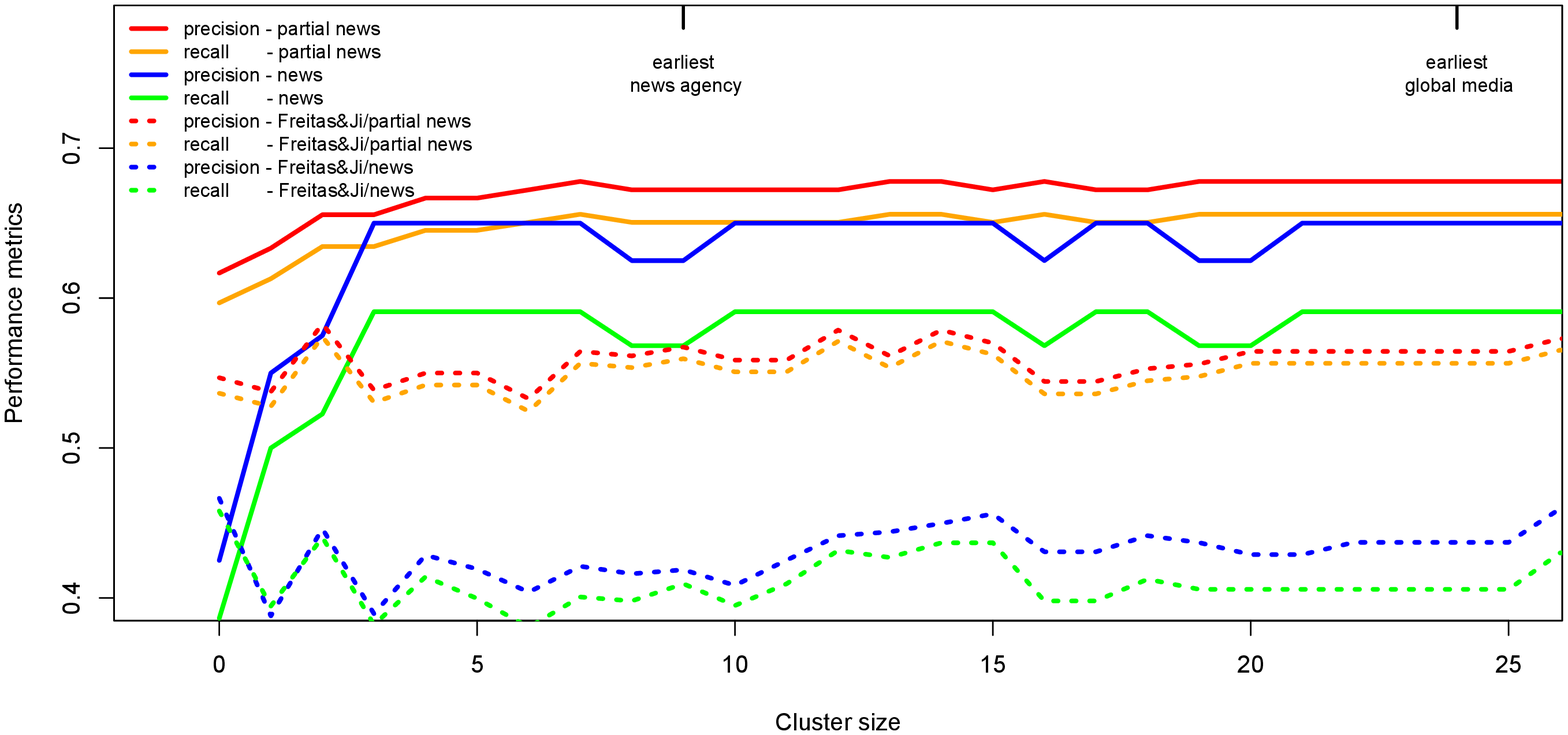}
\end{subfigure}
\caption{Performance of newsworthiness ranking. Top: PDF of newsworthiness scores (rescaled to $[0,1]$) on 400 sampled clusters. Thresholds for news, partial news, and non-news contents are displayed. Miniature graphic shows how newsworthiness thresholds were determined by F-measure; Bottom: precision and recall for news and partial news detection as clusters grow, benchmarked against the active learning model proposed by Freitas and Ji \cite{freitas2016identifying}.}
\label{fig:news_pred}
\end{figure}

Reuters journalists advised us to set up a three-grade criteria of judging newsworthiness for an event: (1) \textit{newsworthy} - events with significant impact and global interests that can be reported by major news agencies like Reuters; (2) \textit{partially newsworthy} - events with local impact that can be reported on local news media; (3) \textit{not newsworthy}. Therefore, the newsworthiness evaluation is considered as a ranking task. Ideally, newsworthiness scores should reflect the order of these three grades and separate them at certain thresholds. To conduct this evaluation, we first examine the distribution of newsworthiness scores generated by our algorithm. Using all clusters created in a week's period starting from 01/21/2017, we plotted the estimated probability density function of newsworthiness scores in Figure \ref{fig:news_pred}. The plot is quite close to a normal distribution. Intuitively, one can image its left and right tails corresponding to a small amount of escaped noise (from filtering algorithms) on the left, and important news on the right. The majority of other clusters in the the middle are a mixture of newsworthy and partially newsworthy stories. We sampled 400 clusters and two annotators evaluated their newsworthiness grades with inter-agreement Kappa 0.68 (we used weighted Kappa that considers order of a multi-grade scale \cite{weightedKappa}) and obtain $11\%$ news and $31\%$ partial news. In terms of ranking, the overall quality of the newsworthiness score in NDCG is 0.84. We then treated the evaluation as two classification tasks of 1) detecting news, and 2) detecting both news and partial news. We experimentally set the thresholds as shown in the top segment of Figure \ref{fig:news_pred}. The results show that our algorithm can detect partial news and news at 0.68 precision \& 0.66 recall, and news-only at 0.67 precision \& 0.59 recall. Since newsworthiness scores are updated as clusters grow, we also checked their dynamic performance as illustrated in the bottom of Figure \ref{fig:news_pred}. Before the event is reported by any news agency, our algorithm can recognize news as partial news with 0.61 precision and recall. But it does not perform as well in capturing news-only events at such an early stage (precision and recall are 0.58 and 0.52, respectively). One reason is that journalists can have different perceptions of the newsworthiness of the same event under different scenarios scenarios. For example, an explosion at New York is more newsworthy than a city in a war-torn area. Thus, it is hard for a topic-based model to tell such a difference. This problem can be tackled by carefully tailored news feeds for particular customers, as described in Section \ref{sc}.    

We benchmarked our ranking model against Freitas and Ji \cite{freitas2016identifying}. They use a SVM model with a polynomial kernel to identify newsworthiness using a set of content-based features. Their best performing model is trained in an active setting: by curating the training data against an earlier model \cite{zubiaga2014} using a \textit{Query-by-Committee} method. We implemented a similar model and used it as a benchmark against \textit{Tracer}. As Figure \ref{fig:news_pred} shows, \textit{Tracer} outperforms the benchmark in both tasks of detecting news and detecting partial news.

\subsection{Veracity}
Evaluating the veracity algorithm in a production system is somewhat different from using experimental data as in \cite{liu2015rumor}. The ratio of false rumors to true stories that appear on \textit{Tracer}'s daily feed is much smaller than the balanced experimental set-up. Even from Aug. to Nov. 2016 when fake news surrounding the US presidential campaigns was rampant, we found only $<0.01\%$ \textit{Tracer} clusters to be about the viral fake news stories reported by BuzzFeed \cite{SilvermanFakeNewsFacebook}. Thus, we approached the veracity evaluation as a fraud detection problem \cite{delamaire2009credit} since it also deals with a skewed data problem. Just as in fraud detection, we evaluate the system on mis-classification cost and false negatives rather than model accuracy. In terms of news, we focus on the precision of veracity scores assigned to stories labeled at least partially newsworthy since we can't afford to surface rumors as news. We also check which percentage of stories labeled as false are actual rumors. Using the same pool of clusters above, we randomly sampled 100 instances of stories labeled newsworthy, 100 partially newsworthy and 100 rumor clusters (veracity score $<0$). Two annotators assessed their veracity via a four-grade criteria: (1) \textit{true} - news verified by trusted news media; (2) \textit{likely true} - not verified yet but comes from highly credible sources such as governments and journalists; (3) \textit{false} - debunked by news media or rumor checkers such as \texttt{snopes.com}; (4) \textit{likely false} - not debunked but very suspicious and likely false. If they cannot make a decision on a cluster, we ignore it and re-sample an additional story. Coder agreement in weighted Kappa is 0.81. To confirm the performance of our system on widespread fake news of the 2016 US presidential election, we used 20 famous fake stories from the BuzzFeed report\footnote{https://goo.gl/eJ2S2z} mentioned above and found 115 \textit{Tracer} clusters corresponding to 16 stories out of 20. We measured the performance of our veracity score and the 5-dot veracity indictor on these four datasets as two binary classification problems (true \& likely true as 1 and false \& likely false as 0 for truth detection; opposite for rumor detection) in Table \ref{tbl:verify_metric}. The result shows that our algorithm can reach $>95\%$ precision for detecting truth when a cluster is just created. When it flags a rumor, with a $>60\%$ chance it is false. We can conclude that \textit{Tracer} can verify true stories reliably and debunk false information with decent accuracy on a routine basis. However, when fake news surges such as in political elections, our system can only flag about $65-75\%$ rumor clusters. Our error analysis reveals the difficulties of creating a reliable rumor detector at an early stage. Verified Twitter users can be fooled and help spread fake news. No matter how good our clustering algorithm is, sometimes true and fake news (election news in our case), can be mixed together unavoidably if they are about very similar topics. Using source credibility will not work in these scenarios. A possible solution is to rely on a fact checking algorithm \cite{wu2014toward}, which will be one of our future investigations.  

\begin{table}
\centering
\caption{Precision (prec.) or recall (rec.) of veracity prediction at early (i.e. when a cluster is just created with 3 tweets) and developing stages (when a cluster grows to 30 tweets) of an event's evolution. A fair judgement uses the 0 score threshold to separate truth from rumors (rumor: score < 0 and truth: score > 0). Strict judgement buffers truth from rumors by a margin (i.e. rumors should fall in the ``red'' and truth in the ``green'' region on the UI). Loose judgement includes the yellow indicator in addition.} \label{tbl:verify_metric}
\setlength\tabcolsep{3pt}
\resizebox{\columnwidth}{!}{
\begin{tabular}{|l|c|c|c|c|c|c|c|c| } \hline  
\multirow{2}{*}{Dataset}   &  True & \multirow{2}{*}{Metric} & \multicolumn{2}{c|}{Fair}  & \multicolumn{2}{c|}{Strict} & \multicolumn{2}{c|}{Loose}  \\ \cline{4-9}
                 &  Ratio & & 3       & 30        & 3       & 30       & 3       & 30                \\ \hline 
Pred. News       & $99\%$ & Prec. & 0.98    & 0.99      & 0.99    & 0.99     & 0.98    & 0.99              \\  
Pred. Part. News & $91\%$ & Prec. & 0.96    & 0.97      & 0.97    & 0.98     & 0.93    & 0.95              \\ \hline \hline
                 &  False & \multirow{2}{*}{Metric} & \multicolumn{2}{c|}{Fair}  & \multicolumn{2}{c|}{Strict } & \multicolumn{2}{c|}{Loose}  \\ \cline{4-9}
                 & Ratio   &  & 3   & 30       & 3      & 30       & 3       & 30                 \\ \hline 
Pred. Rumors     & $62\%$  & Prec.  & 0.63   & 0.62      & 0.63   & 0.63     & 0.56    & 0.59               \\
Fake News        & $100\%$ & Rec.   & 0.61   & 0.64      & 0.54   & 0.57     & 0.73    & 0.76               \\ \hline
\end{tabular}
}
\end{table}

\begin{figure}
\centering
\includegraphics[width=0.5\textwidth, height=0.24\textheight]{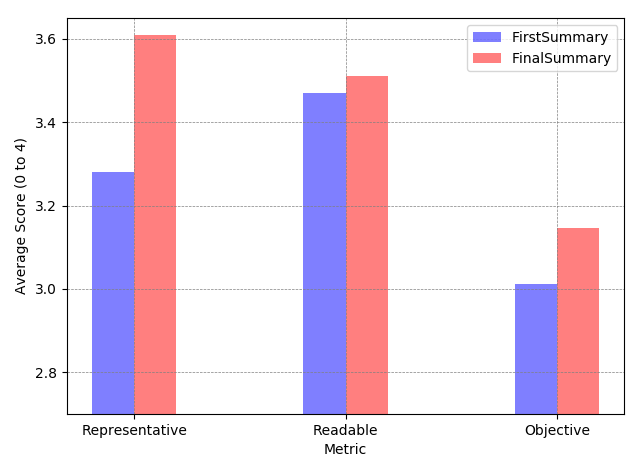}
\caption{Performance of the summarization algorithm on early and final stages of clustering.}
\label{fig:summarization}
\end{figure}

\subsection{Summary - Headline}
 A summary is generated as soon a cluster is formed with three tweets. After that, whenever there is a merge, \textit{Tracer} tries to re-generate a summary. In this experiment, we use human evaluators to assess the quality of the summary. We selected 100 event clusters, and generated two summaries for each event: when a cluster is initially generated using three tweets, the summary assigned to it is labeled ``FirstSummary;'' after the cluster becomes idle and no longer grows, the summary of the cluster at that point is labeled ``FinalSummary.'' The goal was to assess the summary quality, and to see if there was any difference between FirstSummary and FinalSummary. Three user-perceived metrics were used in the evaluation: \textit{representative, readable} and \textit{objective} \cite{alsaedi2016summary}. \textit{Representative} reflects how well the summary represents the event. \textit{Readable} refers to its readability. \textit{Objective} shows if the tweet objectively describes an event or reflects personal opinion. These three dimensions ensure that opinionated, informal, illegible or uninformative tweets are not selected as the summary of an event. 
 
A 5-point scale, from 0 (lowest) to 4 (highest), was used for evaluation. Each summary was evaluated by two annotators. For a given cluster, the annotators went through all its tweets, and assigned a score for each metric (representative, readable, objective) of each of the two summaries (FirstSummary and FinalSummary). Figure \ref{fig:summarization} shows the average scores for the 100 events. It shows that both summaries perform very well on all three metrics, with a score greater than 3. For \textit{representativeness}, FinalSummary is much better than FirstSummary, and the difference is statistically significant at \textit{p=0.01} using \textit{t-test}. This reflects the necessity of regenerating the summary when a cluster is updated. The weighted Kappa values for \textit{representative, readable}, and \textit{objective} of FirstSummary and FinalSummary are 0.75, 0.87, 0.79, 0.77, 0.88 and 0.79, respectively. The result also indicates even first summaries are good enough (all scores > 3.0) for news headlines.

\subsection{Timeliness}\label{timeliness}
News business is largely reliant on the timely detection of important events. In order to evaluate \textit{Tracer}'s effectiveness as an automated news delivery tool, we analyzed its timeliness compared to mainstream news media\footnote{An earlier version of this study is reported in \cite{liu2016reuters}}. Using Google's Top Stories section \footnote{https://news.google.com} as well as input from Reuters journalists, we identified 66 high-profile news stories between 12/02/2015 and 02/03/2017 \footnote{Data available at https://tinyurl.com/jt3n5og}. 

For each story, we identified the following: 1) The exact time of the event (if we couldn't verify the time of the event, we used the timestamp of the earliest tweet about the event). 2) The timestamp of the first tweet that \textit{Tracer} captured about the event. 3) The nature of the first tweet (i.e. whether it was a witness account, a local reporter, national news media, etc.). 4) The timestamp of the first cluster reported by \textit{Tracer}. 5) The timestamp of the earliest story published by any of three global mainstream news agencies (\texttt{Reuters}, \texttt{AP} and \texttt{AFP}), and three global news media (\texttt{BBC}, \texttt{CNN}, \texttt{Bloomberg}). If unable to verify the timestamp of the earliest story, we used the timestamp of earliest tweet posted by the official handles of these news outlets. 

We divided the 66 events into various subsets. For instance, some events were unexpected (such as terror attacks) while some were expected to happen (e.g. NY state primary results were expected to be announced). The events spanned domestic (i.e. US-based) and international stories. Since \textit{Tracer}'s algorithms were all trained on English-language data, we were also curious to see how well the system performed when reporting events from non-English speaking countries.  As was expected, the system had the highest lead against official news media when an event was unexpected (27 mins) or had taken place outside of the U.S. (28 mins). When it came to the source of news, local authorities (such as fire and police departments, local government agencies and local news media) gave \textit{Tracer} its largest lead (53 mins). For instance, the San Bernardino shooting attack was captured by a tweet posted by the San Bernardino Fire Department.

\begin{figure}
\centering
\includegraphics[width=0.48\textwidth]{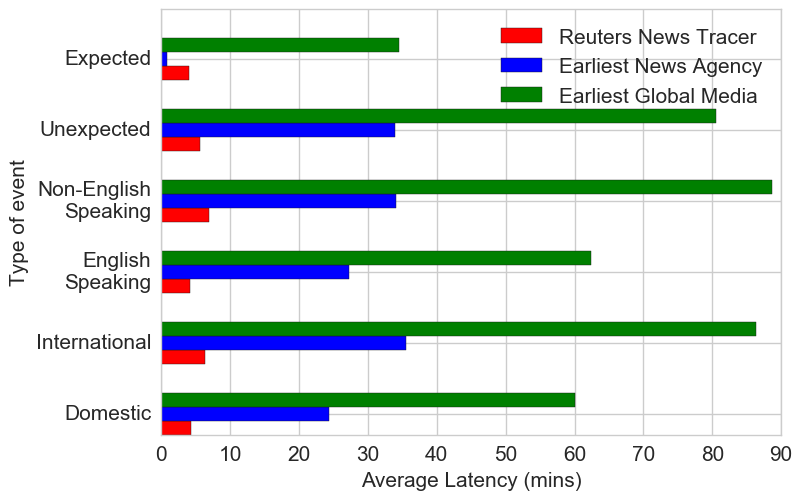}
\caption{Average latency of \textit{Tracer} compared to other media. Tracer consistently leads, except for expected events.} 
\label{fig:latency}
\end{figure}

Figure \ref{fig:latency} shows the result of the comparative timeliness analysis. For each category of news, we calculated \textit{Tracer}'s average latency in detecting the events. We compared this latency to that of  global news agencies and global news media mentioned above. As the figure shows, \textit{Tracer} is almost always ahead of mainstream media, except in the case of expected news. It's important to note that timeliness alone is not a sufficient measure of efficiency for an automated news delivery system. Information load and screen persistence are just as important as timeliness. \textit{Tracer} also proved effective when a screen persistence analysis was performed (see \cite{liu2016reuters}). The timeliness study further establishes the effectiveness of \textit{Tracer}'s algorithms in detecting and delivering news. Despite using only 2\% of Twitter's data stream, the tool is able to beat mainstream media in many unexpected, high-profile stories.

\subsection{Fully Automated New Feeds}\label{sceval}
\subsubsection{Performance}
Two Reuters News editors evaluated the \textbf{Disaster Feed} for its ability to report breaking disasters around the world. Due to the highly ambiguous nature of what counts as ``breaking,'' the editors proposed a subjective metric for evaluation. For each headline, they identified whether or not they would choose to compose a story for it. The original business requirement set the alerting ratio at 20\% (i.e. 1 in 5 headlines produced by the feed were required to be ``alertable''). The evaluation surpassed this requirement and set the performance at 50\% (i.e. 1 in every 2 headlines were alertable). The \textbf{Supply Chain Events Feed} was tested on a dataset which was deliberately designed to focus on a non-English speaking country. Two news experts manually curated 74 events that had taken place in Philippines between Aug and Sep 2016. The events ranged from terror attacks, structural fires and social unrest to earthquakes, volcanoes, floods, heat waves, storms, and other extreme weather conditions. Of the 74 events, the system was able to correctly identify 50. In addition, the system was able to find 12 new events that the human curators had failed to find. This proved \textit{Tracer}'s ability to leverage local sources to detect events that would otherwise go unnoticed. Table \ref{tab:sc} summarizes the result of evaluations for the two custom feeds. 

\begin{table}[]
\centering
\caption{The performance of customized disaster and supply chain event detection feeds produced by \textit{Tracer}.}
\label{tab:sc}
\resizebox{\columnwidth}{!}{%
\begin{tabular}{l|c|l|c}
                                                                                        & \textbf{\begin{tabular}[c]{@{}l@{}}Stories per day\end{tabular}} & \textbf{Metric} & \textbf{\begin{tabular}[c]{@{}l@{}}Performance\end{tabular}} \\ \hline
\textbf{\begin{tabular}[c]{@{}l@{}}Disaster\end{tabular}}                        & 101                                                                          & Alertability           & 50\%                                                          \\ \hline
\multirow{2}{*}{\textbf{\begin{tabular}[c]{@{}l@{}}Supply\\ Chain\end{tabular}}} & \multirow{2}{*}{430}                                                         & Precision              & 94.3\%                                                              \\ \cline{3-4} 
                                                                                        &                                                                              & Recall               & 67.5\%                                                               
\end{tabular}
}
\end{table}

\subsubsection{Timeliness}
Table \ref{fig:events} shows three unexpected events that made headlines during October, 2017. The table shows the earliest cluster published by \textit{Tracer}, as well as the earliest alert published to the disaster feed and the earliest news alert by \textit{Reuters}. Due to its design, \textit{Tracer} is often able to detect breaking stories by identifying early witness accounts. This was the case for the mass shooting in Las Vegas. A witness reported the incident at 1:22 AM, which triggered a \textit{Tracer} cluster. The disaster feed is required to only report stories with a high veracity score. The veracity score of the \textit{Tracer} cluster (indicated by four green dots), did not meet the feed's strict requirement. As a result, the feed was updated only after the veracity score reached acceptable level. This happened around 1:39 AM. \textit{Reuters} reported the incident at 1:49 AM.

In the case of the murder of Daphne Caruana Galizia, the first cluster detected by \textit{Tracer} had a high veracity score, and the story was immediately reported to the disaster feed. \textit{Reuters} reported the story almost an hour later.

In the case of the terror attack in New York City, the story was again captured via eyewitness accounts at 3:22 PM. Within three minutes, the veracity score was boosted enough for the story to be published to the disaster feed (where the veracity score is identified as ``confidence score''). \textit{Reuters} followed up a few minutes later.

Timeliness is of utmost important to journalists, and therefore, \textit{Tracer} and its accompanying disaster feed have proven a useful source of breaking news. \textit{Tracer} is often used as a tipping tool, where journalists can discover early signals. The disaster feed on the other hand waits for the machine-generated veracity score to reach a certain level before reporting a story. This slows down the feed, often by a few minutes. Nevertheless the feed is still often ahead of the earliest news alert.

\begin{table*}[t]
\centering
\caption{Three recent events and their corresponding \textit{Tracer}'s and \textit{Reuters} alerts.}
\label{fig:events}
\resizebox{\textwidth}{!}{%
\begin{tabular}{@{}llll@{}}
\toprule
\textbf{Event}                        & \begin{tabular}[c]{@{}l@{}}Mass shooting at Mandalay Bay, Las Vegas\\ Oct 2, 2017\end{tabular} & \begin{tabular}[c]{@{}l@{}}Murder of famed Maltese journalism\\ Oct 16, 2017\end{tabular} & \begin{tabular}[c]{@{}l@{}}Terror attack in NYC\\ Oct 31. 2017\end{tabular} \\ \midrule
\textbf{\begin{tabular}[c]{@{}l@{}}Earliest\\ Tracer\\ cluster\end{tabular}}     &  \raisebox{-.5\height}{\includegraphics[width=2in]{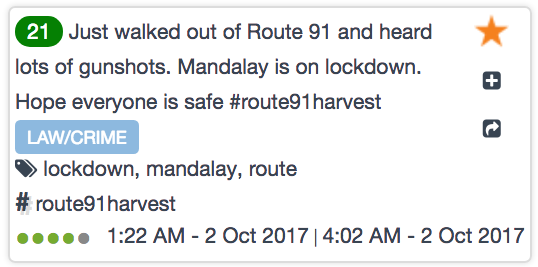}}                                                                                              &     \raisebox{-.5\height}{\includegraphics[width=2in]{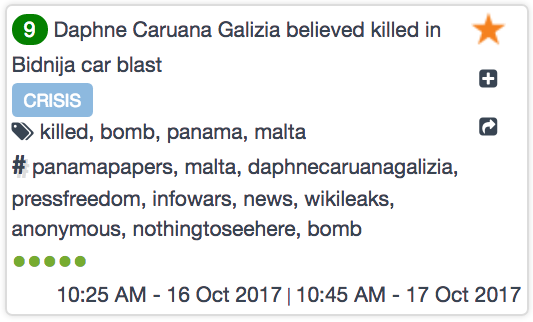}}                                                                                      &        \raisebox{-.5\height}{\includegraphics[width=2in]{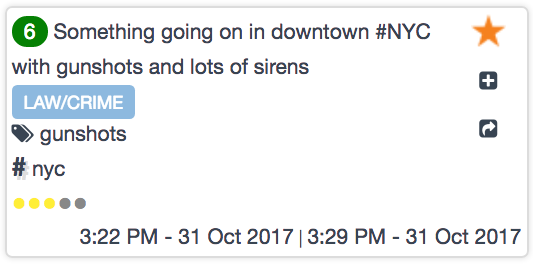}}                                                                     \\ \hline
\textbf{\begin{tabular}[c]{@{}l@{}}Earliest\\ disaster feed\\ alert\end{tabular}} &  \raisebox{-.5\height}{\includegraphics[width=2.5in]{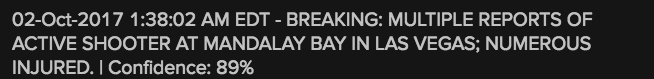}}                                                                                              &         \raisebox{-.5\height}{\includegraphics[width=2.5in]{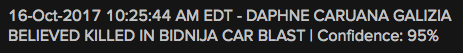}}                                                                                  &      \raisebox{-.5\height}{\includegraphics[width=2.5in]{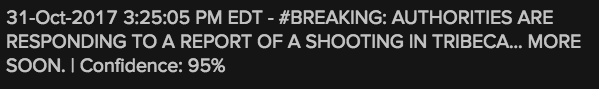}}                                                                       \\ \hline
\textbf{\begin{tabular}[c]{@{}l@{}}Earliest\\ Reuters\\ alert\end{tabular}}      &     \raisebox{-.5\height}{\includegraphics[width=2.5in]{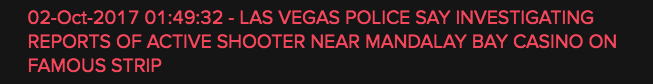}}                                                                                           &       \raisebox{-.5\height}{\includegraphics[width=2.5in]{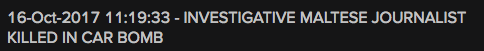}}                                                                                    &       \raisebox{-.5\height}{\includegraphics[width=2.5in]{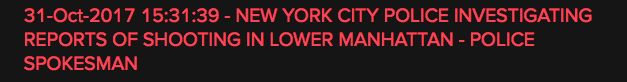}}                                                                      \\ \bottomrule
\end{tabular}}
\end{table*}

\section{Tracer's Successes \& Limitations}

In this paper, we presented \textit{Tracer}, a system designed to achieve fully automated news production. It is now used by Reuters journalists and will be available to Reuters' customers soon. From a data science perspective, the main success of \textit{Tracer} is its design process and its capability of mining critical information from big data. First, we confirmed that Twitter indeed surfaced events earlier than news media. Then we investigated ways to detect them accurately and swiftly, which proved to be a challenging task. We achieved this by sifting through the Twitter stream iteratively, from finding a manageable number of events, to determining their newsworthiness and veracity, to generating news headlines, and finally composing highly important and accurate news feeds. Each of these stages are supported by a suite of algorithms, which employ a mixture of rule-based and learning-based models. Learning models make \textit{Tracer} generalizable to unobserved data, and maintain the system's recall. The customized news feeds carefully select events \& stories with near perfect precision to meet the requirements of users. These pieces working together enable \textit{Tracer} to automate the entire news production pipeline. However, we did not explore algorithmic storytelling \cite{colford2014leap} in depth, since headlines were sufficient for generating news feeds. It will be interesting to explore this idea and determine whether algorithms can write news stories automatically. 

To journalists, the most appealing feature of \textit{Tracer} is its capability to detect news faster than news professionals and its ability to provide some indication of veracity. Our analysis shows that \textit{Tracer} is good at detecting unexpected events such as accidents, conflicts and natural disasters. Even though \textit{Tracer} only supports English, it can still lead mainstream media in reporting from non-English speaking countries. We believe this is due to the fact that many eye witnesses prefer to use English to report major events, in order to notify global media. However, \textit{Tracer} is not ahead on all cases. Journalists can beat \textit{Tracer} in reporting expected news that they are closely following. For example, \textit{Tracer} cannot be expected to know the outcome of elections before news media. Our rumor debunking algorithm also requires further improvement, especially in the rapidly polarizing landscape of political opinions. Our system currently supports two end-to-end feeds, which are carefully crafted and polished by tuning system parameters. Providing a user-driven platform that allows end users to set their own parameters would be a much more flexible solution to build \textit{Tracer}'s feeds according to customer needs. As a result, our journey on news automation is not yet over. There is plenty of room for us to continue improving \textit{Tracer} in the future.

\bibliographystyle{IEEEtran}
\bibliography{AutoNews}

\begin{thebibliography}{10}
\providecommand{\url}[1]{#1}
\csname url@samestyle\endcsname
\providecommand{\newblock}{\relax}
\providecommand{\bibinfo}[2]{#2}
\providecommand{\BIBentrySTDinterwordspacing}{\spaceskip=0pt\relax}
\providecommand{\BIBentryALTinterwordstretchfactor}{4}
\providecommand{\BIBentryALTinterwordspacing}{\spaceskip=\fontdimen2\font plus
\BIBentryALTinterwordstretchfactor\fontdimen3\font minus
  \fontdimen4\font\relax}
\providecommand{\BIBforeignlanguage}[2]{{%
\expandafter\ifx\csname l@#1\endcsname\relax
\typeout{** WARNING: IEEEtran.bst: No hyphenation pattern has been}%
\typeout{** loaded for the language `#1'. Using the pattern for}%
\typeout{** the default language instead.}%
\else
\language=\csname l@#1\endcsname
\fi
#2}}
\providecommand{\BIBdecl}{\relax}
\BIBdecl

\bibitem{czarniawska2011cyberfactories}
B.~Czarniawska-Joerges, \emph{Cyberfactories: How news agencies produce
  news}.\hskip 1em plus 0.5em minus 0.4em\relax E. Elgar Publishing, 2011.

\bibitem{wang2015nyt}
S.~Wang, ``The new york times built a slack bot to help decide which stories to
  post to social media,'' \emph{NiemanLab}, 2015.

\bibitem{colford2014leap}
P.~Colford, ``A leap forward in quarterly earnings stories,'' \emph{The
  definitive source - AP blog}, 2014.

\bibitem{Stray:2016}
J.~Stray, ``The age of the cyborg,'' \emph{Columbia Journalism Review}, 2016.

\bibitem{Al-Kofahi:2017}
\BIBentryALTinterwordspacing
K.~Al-Kofahi. (2017) Becoming smarter about credible news. [Online]. Available:
  \url{https://www.oreilly.com/ideas/becoming-smarter-about-credible-news}
\BIBentrySTDinterwordspacing

\bibitem{diakopoulos2016algorithmic}
N.~Diakopoulos and M.~Koliska, ``Algorithmic transparency in the news media,''
  \emph{Digital Journalism}, pp. 1--20, 2016.

\bibitem{schifferes2014identifying}
S.~Schifferes, N.~Newman, N.~Thurman, D.~Corney, A.~G{\"o}ker, and C.~Martin,
  ``Identifying and verifying news through social media: Developing a
  user-centred tool for professional journalists,'' \emph{Digital Journalism},
  vol.~2, no.~3, pp. 406--418, 2014.

\bibitem{marcus2011twitinfo}
A.~Marcus, M.~S. Bernstein, O.~Badar, D.~R. Karger, S.~Madden, and R.~C.
  Miller, ``Twitinfo: aggregating and visualizing microblogs for event
  exploration,'' in \emph{SIGCHI}.\hskip 1em plus 0.5em minus 0.4em\relax ACM,
  2011, pp. 227--236.

\bibitem{graefe2016guide}
A.~Graefe, ``Guide to automated journalism,'' \emph{Tow Center for Digital
  Journalism. Janeiro}, 2016.

\bibitem{Sankaranarayanan:2009}
J.~Sankaranarayanan, H.~Samet, B.~E. Teitler, M.~D. Lieberman, and J.~Sperling,
  ``Twitterstand: News in tweets,'' in \emph{SIGSPATIAL}.\hskip 1em plus 0.5em
  minus 0.4em\relax New York, NY, USA: ACM, 2009, pp. 42--51.

\bibitem{osborne2014facebook}
M.~Osborne and M.~Dredze, ``Facebook, twitter and google plus for breaking
  news: Is there a winner?'' in \emph{ICWSM}, 2014.

\bibitem{liu2016reuters}
X.~Liu, Q.~Li, A.~Nourbakhsh, R.~Fang, M.~Thomas, K.~Anderson, R.~Kociuba,
  M.~Vedder, S.~Pomerville, R.~Wudali \emph{et~al.}, ``Reuters tracer: A large
  scale system of detecting \& verifying real-time news events from twitter,''
  in \emph{25th ACM International on Conference on Information and Knowledge
  Management}.\hskip 1em plus 0.5em minus 0.4em\relax ACM, 2016, pp. 207--216.

\bibitem{Atefeh2013}
F.~Atefeh and W.~Khreich, ``A survey of techniques for event detection in
  twitter,'' \emph{Computational Intelligence}, 2013.

\bibitem{Yin:2013}
J.~Yin, ``Clustering microtext streams for event identification,'' in
  \emph{IJCNLP}.\hskip 1em plus 0.5em minus 0.4em\relax Nagoya, Japan: Asian
  Federation of Natural Language Processing, October 2013, pp. 719--725.

\bibitem{petrovic2010streaming}
S.~Petrovi{\'c}, M.~Osborne, and V.~Lavrenko, ``Streaming first story detection
  with application to twitter,'' in \emph{HLT}, 2010, pp. 181--189.

\bibitem{li2016tweet}
Q.~Li, S.~Shah, X.~Liu, A.~Nourbakhsh, and R.~Fang, ``Tweet topic
  classification using distributed language representations,'' in
  \emph{IEEE/WIC/ACM International Conference on Web Intelligence}.\hskip 1em
  plus 0.5em minus 0.4em\relax IEEE, 2016, pp. 81--88.

\bibitem{chen2010semantic}
G.~Chen, J.~Warren, and P.~Riddle, ``Semantic space models for classification
  of consumer webpages on metadata attributes,'' \emph{Journal of Biomedical
  Informatics}, vol.~43, no.~5, pp. 725--735, 2010.

\bibitem{harcup2001news}
T.~Harcup and D.~O'neill, ``What is news? galtung and ruge revisited,''
  \emph{Journalism studies}, vol.~2, no.~2, pp. 261--280, 2001.

\bibitem{freitas2016identifying}
J.~Freitas and H.~Ji, ``Identifying news from tweets,'' \emph{NLP+ CSS 2016},
  p.~11, 2016.

\bibitem{Leskovec2009meme}
J.~Leskovec, L.~Backstrom, and J.~Kleinberg, ``Meme-tracking and the dynamics
  of the news cycle,'' in \emph{Proc. of the 15th ACM SIGKDD}, 2009, pp.
  497--506.

\bibitem{khan2009survey}
S.~S. Khan and M.~G. Madden, ``A survey of recent trends in one class
  classification,'' in \emph{Irish Conference on AICS}.\hskip 1em plus 0.5em
  minus 0.4em\relax Springer, 2009, pp. 188--197.

\bibitem{griffiths2004finding}
T.~L. Griffiths and M.~Steyvers, ``Finding scientific topics,'' \emph{Proc. of
  NAS}, vol. 101, no. suppl 1, pp. 5228--5235, 2004.

\bibitem{SilvermanFakeNewsFacebook}
C.~Silverman, ``This analysis shows how viral fake election news stories
  outperformed real news on facebook,'' 2016.

\bibitem{liu2015rumor}
X.~Liu, A.~Nourbakhsh, Q.~Li, R.~Fang, and S.~Shah, ``Real-time rumor debunking
  on twitter,'' in \emph{24th ACM International on Conference on Information
  and Knowledge Management}, 2015, pp. 1867--1870.

\bibitem{castillo2011information}
C.~Castillo, M.~Mendoza, and B.~Poblete, ``Information credibility on
  twitter,'' in \emph{WWW}, 2011, pp. 675--684.

\bibitem{nourbakhsh2015rumor}
A.~Nourbakhsh, X.~Liu, S.~Shah, R.~Fang, M.~M. Ghassemi, and Q.~Li,
  ``Newsworthy rumor events: A case study of twitter,'' in \emph{2015 IEEE
  International Conference on Data Mining Workshop}, 2015, pp. 27--32.

\bibitem{silverman2014verification}
C.~Silverman, \emph{Verification handbook}, 2014.

\bibitem{zhao2015enquiring}
Z.~Zhao, P.~Resnick, and Q.~Mei, ``Enquiring minds: Early detection of rumors
  in social media from enquiry posts,'' in \emph{WWW}.\hskip 1em plus 0.5em
  minus 0.4em\relax ACM, 2015, pp. 1395--1405.

\bibitem{erkan2004lexrank}
G.~Erkan and D.~R. Radev, ``Lexrank: Graph-based lexical centrality as salience
  in text summarization,'' \emph{JAIR}, vol.~22, pp. 457--479, 2004.

\bibitem{becker2011selecting}
H.~Becker, M.~Naaman, and L.~Gravano, ``Selecting quality twitter content for
  events.'' \emph{ICWSM}, vol.~11, 2011.

\bibitem{li2017novelty}
Q.~Li, A.~Nourbakhsh, S.~Shah, and X.~Liu, ``Real-time novel event detection
  from social media,'' in \emph{IEEE 33rd International Conference on Data
  Engineering}.\hskip 1em plus 0.5em minus 0.4em\relax IEEE, 2017, pp.
  1129--1139.

\bibitem{nourbakhsh2017disasters}
A.~Nourbakhsh, Q.~Li, X.~Liu, and S.~Shah, ``Breaking disasters: Predicting and
  characterizing the global news value of natural and man-made disasters,'' in
  \emph{KDD workshop on Data Science+Journalism}, 2017.

\bibitem{Ozdikis:2016}
O.~Ozdikis, H.~O{\u{g}}uzt{\"u}z{\"u}n, and P.~Karagoz, ``A survey on location
  estimation techniques for events detected in twitter,'' \emph{Knowledge and
  Information Systems}, pp. 1--49, 2016.

\bibitem{Mandl2008}
T.~Mandl, F.~Gey, G.~Di~Nunzio, N.~Ferro, R.~Larson, M.~Sanderson, D.~Santos,
  C.~Womser-Hacker, and X.~Xie, \emph{GeoCLEF 2007: The CLEF 2007
  Cross-Language Geographic Information Retrieval Track Overview}.\hskip 1em
  plus 0.5em minus 0.4em\relax Springer, 2008, pp. 745--772.

\bibitem{Macdonald:2013}
C.~Macdonald, R.~McCreadie, M.~Osborne, I.~Ounis, S.~Petrovic, and
  L.~Shrimpton, ``Can twitter replace newswire for breaking news,'' in
  \emph{ICWSM}, 2013.

\bibitem{bci:2016}
B.~C. Institute, ``Supply chain resilience report,'' 2016.

\bibitem{weightedKappa}
J.~Cohen, ``Weighed kappa: Nominal scale agreement with provision for scaled
  disagreement or partial credit,'' \emph{Psychological Bulletin}, vol.~70,
  no.~4, pp. 213--220, 1968.

\bibitem{zubiaga2014}
A.~Zubiaga and H.~Ji, ``Tweet, but verify: Epistemic study of information
  verification on twitter,'' \emph{Social Network Analysis and Mining}, 2014.

\bibitem{delamaire2009credit}
L.~Delamaire, H.~Abdou, and J.~Pointon, ``Credit card fraud and detection
  techniques: a review,'' \emph{Banks and Bank systems}, vol.~4, no.~2, pp.
  57--68, 2009.

\bibitem{wu2014toward}
Y.~Wu, P.~K. Agarwal, C.~Li, J.~Yang, and C.~Yu, ``Toward computational
  fact-checking,'' \emph{Proceedings of the VLDB Endowment}, vol.~7, no.~7, pp.
  589--600, 2014.

\bibitem{alsaedi2016summary}
N.~Alsaedi, P.~Burnap, and O.~Rana, ``Automatic summarization of real world
  events using twitter.'' \emph{ICWSM}, 2016.

\end{thebibliography}

\end{document}